\documentclass[twocolumn,prl]{revtex4}
\usepackage{amssymb}
\usepackage{graphicx}

\input{tcilatex}

\begin{document}

\title{Isotope engineering in carbon nanotube systems}
\author{F. Simon}
\author{Ch. Kramberger}
\author{R. Pfeiffer}
\author{H. Kuzmany}
\address{Institut f\"{u}r Materialphysik, Universit\"{a}t Wien, Strudlhofgasse 4,
A-1090 Wien, Austria}
\author{V. Z\'{o}lyomi}
\author{J. K\"{u}rti}
\address{Department of Biological Physics, E\"{o}tv\"{o}s University, Budapest,
Hungary}
\author{P. M. Singer}
\author{H. Alloul}
\address{Laboratoire de Physique des Solides, UMR 8502, Universit\'{e} Paris-Sud, B%
\^{a}timent 510, 91405 Orsay, France}

\begin{abstract}
We report on single-wall carbon nanotube (SWCNT) specific $^{13}$C isotope
enrichment. The high temperature annealing of isotope enriched fullerenes
encapsulated in SWCNTs yields double-wall carbon nanotubes (DWCNTs) with a
high isotope enrichment of the inner wall. The vibrational spectra evidences
that no carbon exchange occurs between the two walls. The method facilitates
the identification of the Raman signal of the outer and inner tubes. Nuclear
magnetic resonance proves the significant contrast of the isotope enriched
SWCNTs as compared to other carbon phases, and provides information on the
electronic properties of the small diameter inner tubes of the DWCNTs.
\end{abstract}

\maketitle


Isotope engineering (IE) of materials provides an important degree of
freedom for both fundamental studies and applications. The change in phonon
energies upon isotope substitution, while leaving the electronic properties
unaffected, have given insight into underlying fundamental mechanisms, such
as phonon-mediated superconductivity \cite{BCS}. Applications of IE involve
for instance the controlled doping of isotope engineered ready-prepared Si
heterostructures by means of neutron irradiation \cite{NTD}, fabrication of
mononuclear devices with controlled heat conducting properties \cite%
{CapinskiAPL}, and the basic architecture for spintronics and quantum
computing \cite{ShlimakCM}. Recently, single-wall carbon nanotubes (SWCNTs)
have been intensively studied as a result of the envisaged broad range of
applicability and their unique physical properties related to their quasi
one-dimensional electronic structure. Examples include the presence of Van
Hove singularities in the electronic density of states \cite{Dresselhaus},
and the Tomonaga-Luttinger liquid behavior \cite{KatauraNAT}. IE of carbon
nanotubes using isotope enriched graphite as the starting material was
attempted in order to allow NMR spectroscopy \cite{TangNMRSCI,GozeBacCAR}.
However, the NMR studies have been hampered by the fact that the $^{13}$C
NMR active nuclei can be found in all species of carbons inevitably present
in the purified SWCNT materials, and no nanotube selective enrichment or
purification could be achieved until now. Vibrational spectroscopic methods
are appropriate choices to study the effect of IE. For SWCNTs, Raman
spectroscopy has proven to be the most convenient to characterize their
electronic and structural properties. In addition, Raman studies on
double-wall carbon nanotubes (DWCNTs) \cite{BandowCPL}, synthesized from
fullerene peapods \cite{SmithNAT}, provides a unique opportunity to study
the small diameter inner tubes, while the outer ones are left intact during
the synthesis.

In this Letter, we demonstrate that the combination of isotope engineered
DWCNTs and Raman spectroscopy can give an unprecedented opportunity to
identify the as yet unobserved vibrational modes, such as the D-modes of
very small diameter SWCNTs. We achieved this by developing a $^{13}$C
isotope enrichment method of SWCNTs specific to nanotubes. We first
synthesized SWCNTs encapsulating $^{13}$C enriched C$_{60}$ and C$_{70}$
fullerene molecules, then a high temperature treatment transformed the
fullerenes into a second smaller diameter inner tube that is isotope
enriched, which reflects the enrichment level of the fullerenes. A
significant contrast with respect to the non-enriched carbon phases is
achieved, as proven by NMR spectroscopy, and allows the direct study of very
small diameter SWCNTs.


Commercial SWCNT material (Nanocarblab, Moscow, Russia \cite{nanocarblab}), $%
^{13}$C isotope enriched fullerenes (MER Corp., Tucson, USA), and fullerenes
of natural carbon (Hoechst AG, Frankfurt, Germany) were used to prepare
fullerene peapods C$_{60}$,C$_{70}$@SWCNT. The SWCNTs were purified by the
supplier to 50 \%. The tube diameter distribution was determined from Raman
spectroscopy \cite{KuzmanyEPJB} and we obtained $d=$ 1.40 nm and $\sigma $ =
0.10 nm for the mean diameter and the variance of the Gaussian distribution,
respectively. We used two supplier provided grades of $^{13}$C enriched
fullerenes: 25 and 89 \% (whose values are slightly refined in our study).
The 25 \% and the 89 \% grades were both composed of C$_{60}$/C$_{70}$%
/higher fullerene mixtures in relative proportions of 75 \%:20 \%:5 \%
(defined as $^{\text{25 \%, 13}}$C$_{60}$ in the following) and 12 \%:88 \%:%
\TEXTsymbol{<}1 \% (defined as $^{\text{89 \%, 13}}$C$_{70}$ in the
following), respectively. The natural carbon C$_{60}$ ($^{\text{Nat}}$C$%
_{60} $) had a purity of \TEXTsymbol{>} 99 \%. The SWCNTs and the fullerenes
were sealed under vacuum in a quartz ampoule and annealed at 650 $^{\text{o}%
} $C for 2 hours for the fullerene encapsulation \cite{KatauraSM}. The
peapods were transformed to DWCNTs by a dynamic vacuum treatment at 1250 $^{%
\text{o}} $C for 2 hours following Ref. \cite{BandowCPL}. Vibrational
analysis was performed on a Dilor xy triple Raman spectrometer in the
1.64-2.54 eV (676-488 nm) energy range at 90 K. The spectral resolution was
0.5-2 cm$^{-1} $ depending on the laser wavelength and the resolution mode
used (high or normal resolution). \textit{Ab initio} calculations were
performed with the Vienna \textit{ab initio} Simulation Package (VASP) \cite%
{KressePRB}. Magic angle spinning (MAS) and static $^{13}$C-NMR spectra were
measured in air at room temperature using a Chemagnets (Varian Inc.) MAS
probe at 7.5 Tesla. The $^{13}$C-NMR spectra were obtained by a Fourier
transformation of the free induction decay following the excitation pulse.

\begin{figure}[tbp]
\includegraphics[width=1.0\hsize]{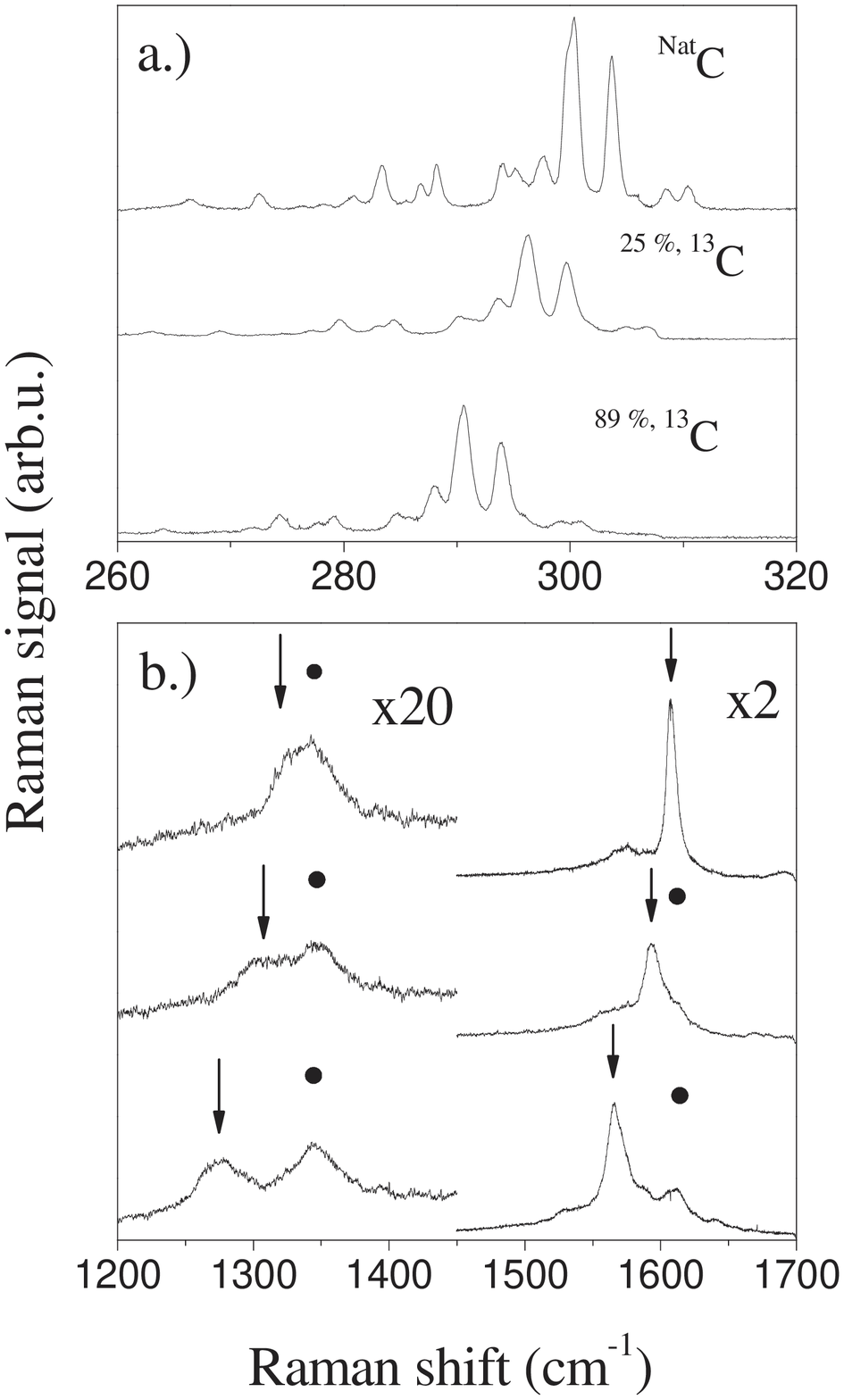}
\caption{Raman spectra of DWCNTs with $^{Nat}$C and $^{13}$C enriched
inner tubes at $\protect\lambda$=676 nm laser excitation and 90 K. a.) RBM
Raman response b.) D and G mode Raman response. Arrows and filled circles
indicate the D and G modes corresponding to the inner and outer tubes,
respectively.}
\label{DWCNT_spectra}
\end{figure}

In Fig. \ref{DWCNT_spectra} we show the Raman spectra of DWCNTs based on $^{%
\text{Nat}}$C$_{60}$, $^{\text{25 \%, 13}}$C$_{60}$ and $^{\text{89 \%, 13}}$%
C$_{70}$ peapods ($^{\text{Nat}}$C-, $^{\text{25 \%, 13}}$C- and $^{\text{89
\%, 13}}$C-DWCNT) for the radial breathing mode (RBM) (Fig. \ref%
{DWCNT_spectra}a), and the D and G mode spectral ranges (Fig. \ref%
{DWCNT_spectra}b) at 676 nm excitation. The narrow lines in Fig. \ref%
{DWCNT_spectra}a were previously identified as the RBMs of the inner tubes 
\cite{PfeifferPRL}. An overall downshift of the inner tube RBMs is observed
for the $^{13}$C enriched materials accompanied by a broadening of the line.
The downshift is clear evidence for the effective $^{13}$C enrichment of
inner tubes. The magnitude of the enrichment and the origin of the
broadening are discussed below.

The RBM lines are well separated for inner and outer tubes due to the $\nu _{%
\text{RBM}}\propto $1/$d$ relation, where $d$ is the tube diameter. However,
other vibrational modes such as the defect induced D and the tangential G
modes strongly overlap for inner and outer tubes. Arrows in Fig. \ref%
{DWCNT_spectra}b indicate a gradually downshifting component of the observed
D and G modes. This component is assigned to the D and G modes of the inner
tubes. The sharper appearance of the inner tube G mode, as compared to the
outer components, is related to the excitation of semiconducting inner tubes
and the Fano line-shape broadened metallic outer tubes \cite%
{PfeifferPRL,SimonCM2}.

The relative magnitude of the inner to outer tube D modes are comparable, as
best seen for the highest enriched sample. The D mode originates from a
double resonance process, which is induced by the defects in the sample \cite%
{ThomsenPRL,KurtiJ20021,ZACHARY20031}. The inner tubes were shown to contain
significantly less defects than the outer ones, as proven by the narrow RBM
phonon line-widths \cite{PfeifferPRL}. Thus, the experimentally observed D
band intensity ratio suggests that a competitive effect, such as the
enhanced electron-phonon coupling in small diameter SWCNTs \cite{BenedictPRB}%
, compensates for the smaller number of defects.

The shift of the RBM, D and G modes were analyzed for the two grades of
enrichment. We found that $\left( \nu _{0}-\nu \right) /\nu _{0}=0.0109(3)$
and $0.0322(3)$ for the 25 and 89 \% samples, respectively. Here, $\nu _{0}$
and $\nu $ are the Raman shifts of the same inner tube mode in the natural
carbon and enriched materials, respectively. In the simplest continuum model
approximation, the shift originates from the increased mass of the inner
tube walls. This gives $\left( \nu _{0}-\nu \right) /\nu _{0}=1-\sqrt{\frac{%
12+c_{0}}{12+c}}$, where $c$ is the concentration of the $^{13}$C enrichment
on the inner tube, and $c_{0}=0.011$ is the natural abundance of $^{13}$C in
carbon. This gives $c=27.7(7)$ and $82.4(8)$ $^{13}$C enrichment for the 25
and 89 \% samples, respectively. The difference between the supplier
provided values and those determined herein reflects the uncertainties in
the enrichment level determination.

We have verified the validity of the continuum approximation for the RBM by
performing first principles calculations on the (5,5) tube as an example.
First, the Hessian was determined by DFT calculations. Then, a large number
of random $^{13}$C distributions were generated and the RBM vibrational
frequencies were determined from the diagonalization of the corresponding
dynamical matrix for each individual distribution. We observed that the
distribution of the resulting RBM frequencies can be approximated by a
Gaussian whose center and variance determines the isotope shifted RBM, $\nu $%
, and a spread in the RBM frequencies. We found the difference between the
shift determined from the continuum model and from the \textit{ab initio}
calculations to be below 1 \%.

\begin{figure}[tbp]
\includegraphics[width=0.9\hsize]{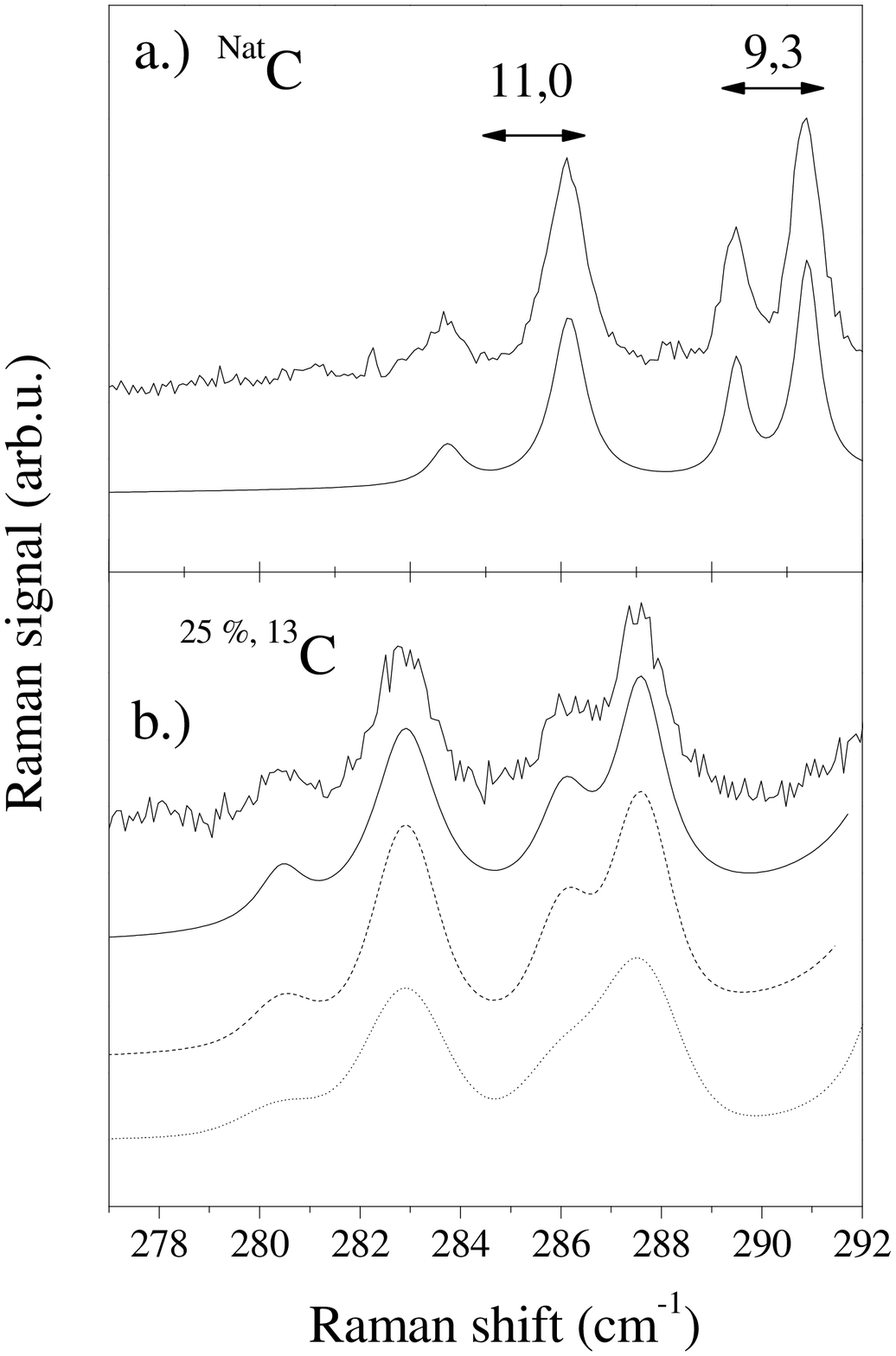}
\caption{RBMs of the (11,0) and the (9,3) inner tubes at $\protect\lambda$%
=676 nm laser excitation with 0.5 cm$^{-1}$ spectral resolution. a.) $^{%
Nat}$C, b.) $^{25 \%, 13}$C. Smooth solid curves are the simulated
Lorentzian components after deconvolution by the spectrometer response.
Dashed curve is simulated line-shape with an extra Gaussian broadening to
the intrinsic lines of the $^{Nat}$C material. Dotted curve is
calculated line-shape (see text).}
\label{broadening}
\end{figure}

The broadening for the $^{13}$C enriched inner tubes is best observed for
the narrow RBMs. In Fig. \ref{broadening}a-b we show the RBMs of the
previously identified (11,0) and (9,3) inner tubes \cite{PfeifferPRL,
KrambergerPRB} with their split components for the $^{\text{Nat}}$C and $^{%
\text{25 \%, 13}}$C samples. Solid curves are the line-shapes after
deconvolution with the Gaussian response of our spectrometer. In Fig. \ref%
{broadening}a, this is a Lorentzian line-shape, but in Fig. \ref{broadening}%
b, the line-shape still contains a Gaussian component, as discussed below.
The FWHMs of the resulting line-shapes are 0.76$(4)$, 0.76$(4)$, 0.44$(4)$,
0.54$(4)$ and 1.28$(6)$, 1.30$(6)$, 1.12$(6)$, 1.16$(6)$ for the 2
components of the (11,0) and (9,3) inner tube RBMs of the $^{\text{Nat}}$C
and $^{\text{25 \%, 13}}$C materials, respectively. The origin of the extra
broadening is due to the random distribution of $^{12}$C and $^{13}$C
nuclei, whose magnitude was determined from the first principles
calculations. We found that the ratio of the half width of extra broadening
and the shift, $\Delta \nu /\left( \nu _{0}-\nu \right) $, is approximately
0.19 for a 30 \% $^{13}$C enriched sample. The corresponding broadened
line-shapes are shown with dotted curves in Fig. \ref{broadening}b. When the
magnitude of the Gaussian randomness related broadening was fit (shown as
dashed curve in Fig.\ref{broadening}b), we found that $\Delta \nu /\left(
\nu _{0}-\nu \right) =0.15$. Similar broadening was observed for the 89 \%
sample which can also be reproduced by the calculation.

\begin{figure}[tbp]
\includegraphics[width=1.0\hsize]{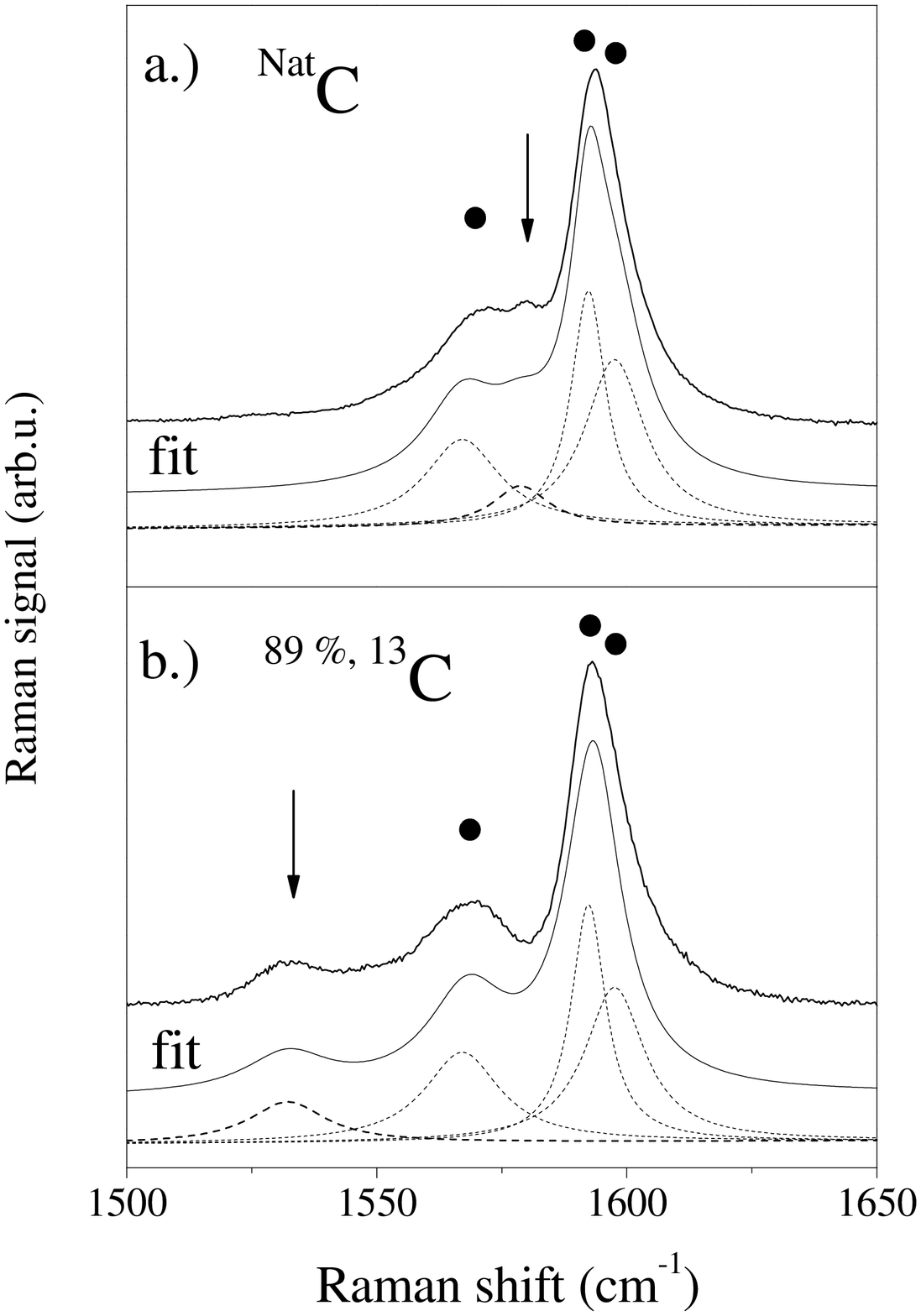}
\caption{G modes of DWCNT with $^{Nat}$C (a) and 89 \% $^{13}$C (b)
enriched inner walls at $\protect\lambda$=488 nm laser excitation and 90 K.
Filled circles indicate the non-shifting components of the outer tube G
modes. Arrows show the only shifting componenent, identified as the inner
tube G mode. Dashed curves show the deconvolution of the observed spectra.}
\label{no_exchange}
\end{figure}

Identification of the different signals also allowed us to address whether
carbon exchange between the inner and outer tubes occurs during the DWCNT
synthesis. In Fig. \ref{no_exchange} we compare the G mode spectra of DWCNTs
based on $^{\text{Nat}}$C$_{60}$ and $^{\text{89 \%, 13}}$C$_{70}$ at 488 nm
excitation. For this excitation, the G mode of the outer tubes dominates the
spectrum since semiconducting outer tubes and conducting inner tubes are in
resonance \cite{SimonCM2}. Indeed, the only shifting component observed is
small (arrows in Fig. \ref{no_exchange}) compared to the non-shifting
components (filled circles in Fig. \ref{no_exchange}). It is known that the
SWCNT G mode consists of several components \cite{PimentaPRB,DubayPRL},
however we deconvoluted the spectra with one Lorentzian line for the shifted
components and three Lorentzians for the non-shifted components. We found
that the spectral region above 1550 cm$^{-1}$ consists of components from
the outer tubes alone, that do not change upon enrichment. Separation of the
inner and outer tube G mode components is more difficult below 1550 cm$^{-1}$%
. We found that within the 1 cm$^{-1}$ experimental accuracy, the outer tube
G modes that are observed above 1550 cm$^{-1}$ are not shifted, giving an
upper limit to the extra $^{13}$C in the outer wall of 1.7 \%.

\begin{figure}[tbp]
\includegraphics[width=0.9\hsize]{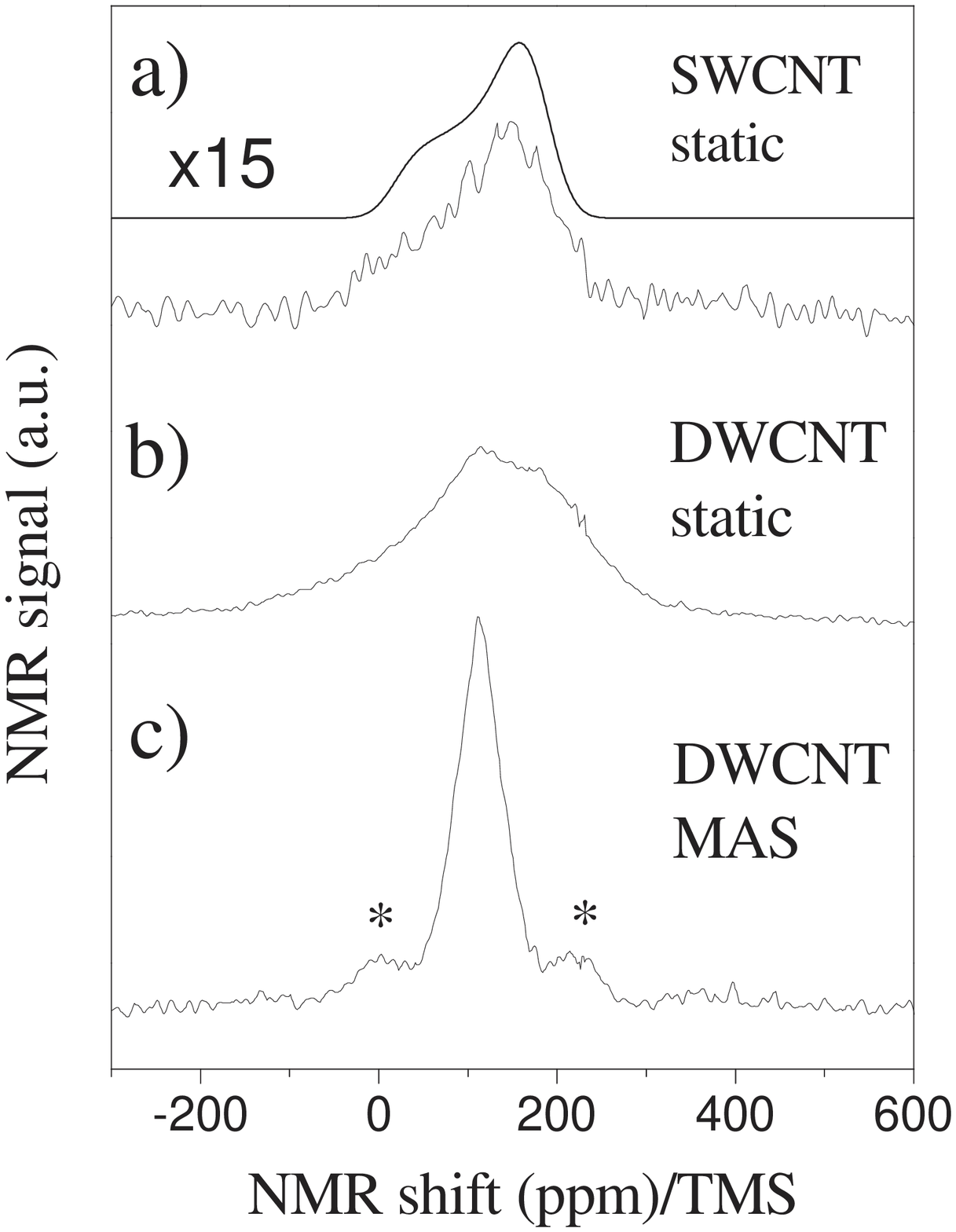}
\caption{NMR spectra normalized by the total sample mass, taken with respect
to the tetramythilesulphate (TMS) shift. (a) Static spectrum for
non-enriched SWNT enlarged by a factor 15. (b) and (c) Static and MAS
spectra of DWCNT based on $^{89 \%, 13}$C$_{70}$ peapods,
respectively. Asterisks show the sidebands at the 8 kHz spinning frequency.
In a) we also show a CSA powder pattern simulation with parameters published
in the literature (smooth solid line) \protect\cite{TangNMRSCI}.}
\label{NMR}
\end{figure}

The significant $^{13}$C isotope enrichment of the inner wall of the DWCNTs
allowed NMR measurements with unprecedented contrast with respect to the
non-enriched other carbon phases. In Fig. \ref{NMR} we show the static and
MAS spectra of $^{13}$C enriched DWCNTs, and the static spectrum for the
SWCNT starting material. The $^{13}$C signal intensity found for the
enriched sample directly monitors the $^{13}$C enrichment of the sample.
This allowed us to determine the carbon fraction belonging to the inner
tubes by comparing the total $^{13}$C-NMR signal intensity of the DWCNT
sample and that of the $^{\text{89 \%, 13}}$C$_{70}$ fullerene starting
material. We found the mass fraction of the inner wall to be 13$($4) \%,
which is consistent with the 15 \% inner tube mass ratio expected from the
SWCNT purity (50 \%), $\sim $70 \% volume filling in highly filled peapod
samples \cite{LiuPRB}, and the mass ratio of encapsulated C$_{60}$ in
SWCNTs. This is the first determination of the inner tube mass content in
DWCNT samples, and proves that DWCNT transformation can be achieved in
macroscopic amounts.

The typical chemical shift anisotropy (CSA) powder pattern is observed for
the SWCNT sample in agreement with previous reports \cite%
{TangNMRSCI,GozeBacCAR}. However, the DWCNT spectra can not be explained
with a simple CSA powder pattern. This is the result of the larger
distribution in chemical shifts since the bonding angle can show a larger
variation among the small diameter inner tubes \cite{KurtiJ20031}. In
addition, the residual line-width in the MAS experiment, which is a measure
of the sample inhomogeneity, is 60(3) ppm, i.e. larger than the $\sim $35
ppm found previously for SWCNT samples \cite{TangNMRSCI,GozeBacCAR}. The
isotropic line position, determined from the MAS measurement, is 111$(2)$
ppm. This value is significantly smaller than the isotropic shift of the
SWCNT samples, 125 ppm \cite{TangNMRSCI,GozeBacCAR}. This difference might
originate from i.)\ the different C-C bonding angle of the small inner
tubes, ii.)\ diamagnetic shielding due to the outer tubes. Currently,
experiments with inner tubes based on different diameter outer ones are
underway to clarify this point.

In conclusion, we report on the synthesis of $^{13}$C enriched SWCNTs. $%
^{13} $C only resides in the SWCNTs, without enriching the inevitable other
carbon phases such as amorphous carbon or graphite. The method is based on
the transformation of $^{13}$C enriched fullerenes encapsulating SWCNTs into
DWCNTs. High levels of isotope enrichment can be achieved without
significant carbon exchange between the two walls. The $^{13}$C enriched
inner tubes facilitate the identification of the vibrational modes of inner
and outer tube components. The SWCNT selective enrichment significantly
simplifies the analysis of NMR experiments where only nanotube sensitive
information is desired. The synthesis method opens the way for the
controllable isotope labelling of SWCNTs without labelling the unwanted
side-products. The described isotope engineering may eventually find
application for the controllable doping of SWCNTs, similar to the isotope
engineering applied in the Si semiconducting industry.

Support from the FWF project Nr. 14893, the OTKA T038014 and the EU projects
BIN2-2001-00580, MEIF-CT-2003-501099, and HPMF-CT-2002-02124 are
acknowledged. Calculations were performed on the Schr\"{o}dinger II cluster
at the University of Vienna.


\begin{thebibliography}{99}
\bibitem{BCS} J. Bardeen, L. N. Cooper, and J. R. Schrieffer, Phys. Rev. 
\textbf{108}, 1175 (1957).

\bibitem{NTD} \textit{Neutron Transmutation Doping of Semiconductors}, ed.
by J. Meese (Plenum Press, NY, 1979).

\bibitem{CapinskiAPL} W. S. Capinski \textit{et al.}, Appl. Phys. Lett. 
\textbf{71}, 2109 (1997).

\bibitem{ShlimakCM} I. Shlimak, cond-mat/0403421.

\bibitem{Dresselhaus} M. S. Dresselhaus, G. Dresselhaus, P. C. Ecklund: 
\textit{Science of Fullerenes and Carbon Nanotubes}, Academic Press, San
Diego 1996.

\bibitem{KatauraNAT} H. Ishii, \textit{et al.}, Nature \textbf{426}, 540
(2003).

\bibitem{TangNMRSCI} X.-P. Tang \textit{et al.}, Science \textbf{288}, 492
(2000).

\bibitem{GozeBacCAR} C. Goze-Bac \textit{et al.}, Carbon \textbf{40}, 1825
(2002).\ 

\bibitem{BandowCPL} S. Bandow \textit{et al.}, Chem. Phys. Lett. \textbf{337}
48 (2001).

\bibitem{SmithNAT} B. W. Smith \textit{et al.}, Nature \textbf{396}, 323
(1998).

\bibitem{nanocarblab} http://www.nanocarblab.com

\bibitem{KuzmanyEPJB} H. Kuzmany \textit{et al.}, Eur. Phys. J. B \textbf{22}
(2001) 307.

\bibitem{KatauraSM} H. Kataura \textit{et al.}, Synth. Met. \textbf{121},
1195 (2001).

\bibitem{KressePRB} G. Kresse and D. Joubert, Phys. Rev. B \textbf{59}, 1758
(1999).

\bibitem{PfeifferPRL} R. Pfeiffer \textit{et al.}, Phys. Rev. Lett. \textbf{%
90} 225501 (2003).

\bibitem{SimonCM2} F. Simon \textit{et al.}, cond-mat/0404110.

\bibitem{ThomsenPRL} C. Thomsen and S. Reich, Phys. Rev. Lett. \textbf{85},
5214 (2000).

\bibitem{KurtiJ20021} J. K\"{u}rti \textit{et al.}, Phys. Rev. B \textbf{65}%
, 165433 (2002).

\bibitem{ZACHARY20031} V. Z\'{o}lyomi \textit{et al.}, Phys. Rev. Lett. 
\textbf{90}, 157401 (2003).

\bibitem{BenedictPRB} L. X. Bendict \textit{et al.}, Phys. Rev. B \textbf{52}%
, 14935 (1995).

\bibitem{KrambergerPRB} Ch. Kramberger \textit{et al.}, Phys. Rev. B \textbf{%
68}, 235404 (2003).

\bibitem{PimentaPRB} M. A. Pimenta \textit{et al.}, Phys. Rev. B \textbf{58}%
, 16016 (1998).

\bibitem{DubayPRL} O. Dubay \textit{et al.}, Phys. Rev. Lett. \textbf{88},
235506 (2002).

\bibitem{LiuPRB} X. Liu \textit{et al.}, Phys. Rev. B \textbf{65} 045419
(2002).

\bibitem{KurtiJ20031} J. K\"{u}rti \textit{et al.}, New. J. Phys. \textbf{5}%
, 125 (2003).
\end{thebibliography}
\end{document}